\documentclass{article}

\usepackage{arxiv}

\usepackage[utf8]{inputenc} 
\usepackage[T1]{fontenc}    
\PassOptionsToPackage{hyphens}{url}\usepackage{hyperref}       
\usepackage{booktabs}       
\usepackage{amsfonts}       
\usepackage{nicefrac}       
\usepackage{microtype}      
\usepackage{lipsum}
\usepackage{tikz}
    \usetikzlibrary{calc}
    \usetikzlibrary{fadings}
\usepackage{upgreek}
    \newcommand{\um}[0]{$\upmu$m} 
\usepackage{xcolor}
    \definecolor{commentgreen}{RGB}{34,139,34}
    \definecolor{stringpurple}{RGB}{160,32,240}
    \definecolor{keywordblue}{RGB}{0,0,255}

\usepackage{listings}
\usepackage{amsbsy}
\usepackage{mathabx}

\usepackage{fancyhdr}
\fancyheadoffset{0pt}
\newcommand{\uns}{\textunderscore}
\renewcommand{\vec}[1]{\boldsymbol{#1}} 

\newcommand{\docVersion}{v.~1.2} 

\title{Gel phantom data for dynamic X-ray tomography}

\author{
  Tommi Heikkil\"{a} \\
  Department of Mathematics and Statistics, \\
  University of Helsinki \\
  Helsinki, Finland \\
  \texttt{tommi.heikkila@helsinki.fi} \\
   \And
  Hanna Help \\
  Department of Physics, \\
  University of Helsinki \\
  \& \\
  Finnish Food Authority \\
  Helsinki, Finland \\
  \texttt{hanna.help@helsinki.fi}
   \And
  Alexander Meaney \\
  Department of Mathematics and Statistics, \\
  University of Helsinki \\
  Helsinki, Finland \\
  \texttt{alexander.meaney@helsinki.fi}
}

\begin{document}
\maketitle

\begin{abstract}
This is the documentation for dynamic X-ray tomography measurements of a gel phantom diffused with potassium iodide contrast agent. The measured data and additional materials are available at \url{http://www.fips.fi/dataset.php} and \href{https://doi.org/10.5281/zenodo.7876521}{Zenodo} for open use by the scientific community, as long as the data and this documentation at \href{https://arxiv.org/}{arXiv} are appropriately referenced.

The files contain: (1) 17 consecutive measurements of the gel phantom organized into \emph{sinograms} at two different resolutions (binning levels) and some additional metadata which can be used to create matrix-free forward operators and filtered back-projection reconstructions; (2) the first measurement using a greater number of projections, and an additional measurement of an 18th time step using very dense angular sampling, for example to be used for reference reconstructions; (3) short example codes to showcase how the data could be used to test and validate reconstruction algorithms.
\end{abstract}


\section{Introduction}
The gel phantom was constructed to simulate diffusion of liquids inside plant stems, namely the flow of iodine-based contrast agents used in high resolution tomographic X-ray imaging of plants. The phantom, which has similar diffusion properties but a higher radiation dose tolerance compared to plant stems, was constructed to test different reconstruction algorithms. It consists of a 50~ml Falcon test tube ($\diameter$ 29~mm $\times$ 115~mm) filled with 2\% agarose gel.

After the agarose solidified, five cavities were made into the gel and filled with 20\% sucrose solution to guarantee the diffusion by directing osmosis to the gel body. In addition, densely punctured plastic straws were placed in the cavities to simulate cellular passages such as \emph{phloem plasmodesmata} and to slow down the lateral diffusion.

The primary measurements consisted of 17 consecutive time frames, with an initial state of no contrast agent followed by steady increase and diffusion into the gel body over time. Each time step consists of 360 projections acquired with a cone-beam micro-CT-scanner. For the data set, we used only the central plane of the cone beam, resulting in 2D fan beam geometry. The measurements and additional files are listed in section \ref{sec:contents}. The measurement setup and the projection geometry are presented in section~\ref{sec:measurements}. Finally, the use of the corresponding forward operators is briefly demonstrated in section~\ref{sec:examples} via an example.

\section{Contents} \label{sec:contents}

The data set contains the following files for MATLAB\footnote{MATLAB is a registered trademark of The MathWorks, inc.}:
\begin{itemize}
    \item \texttt{GelPhantomData\uns b2.mat}
    \item \texttt{GelPhantomData\uns b4.mat}
    \item \texttt{GelPhantom\uns extra\uns frames.mat}
    \item \texttt{gel\uns phantom\uns example.m}
    \item \texttt{gel\uns phantom\uns example\uns GPU.m}
\end{itemize}

The proper use of these files requires \href{https://www.astra-toolbox.com/}{the ASTRA Toolbox} \cite{van2015astra, van2016fast}, \href{https://www.cs.ubc.ca/labs/scl/spot/}{the Spot Linear-Operator Toolbox} \cite{spot}, and \href{https://se.mathworks.com/matlabcentral/fileexchange/74417-heltomo-university-of-helsinki-ct-data-toolbox}{the HelTomo Toolbox} \cite{heltomo}.

\subsection{Sinograms}
 The files \texttt{GelPhantomData\uns b2.mat} and \texttt{GelPhantomData\uns b4.mat} contain a MATLAB structure array which includes for every individual measurement all the necessary metadata and information about the measurement process and geometry, and the \texttt{sinogram}: a numerical $P \times D$ array containing all of the measurements from $P = 360$ projections and $D = 564$ or $256$ detector elements, depending on the detector binning.
 
For computational efficiency the measurements have been binned by a factor of 2 and 4, respectively, before being organized into sinograms. The sinograms are obtained by considering the logarithm of the ratio between the initial intensity and the measured intensity. Each row corresponds to a single projection direction, while each column corresponds to a single detector element. The initial intensity for each individual image is calculated by taking the mean of a $96 \times 96$ pixel area known to be outside the object's X-ray projection.

Binning is a method of lowering the data resolution and improving the signal-to-noise ratio by summing the measured intensity values from multiple neighbouring detector elements, such as non-overlapping $2\times 2$ neighbourhoods. This is also taken into account with the parameters such as relative pixel size. Binning was applied to the raw projection images both horizontally and vertically to conserve the square shape of the detector elements.

If a smaller number of projections is required, for example for sparse angle tomography, the sinograms can be downsampled using the function \texttt{subsample\uns sinogram.m}. This will produce a new tomographic measurement data structure which matches the projection directions given by the user.

\paragraph{Sinogram arrangement:}{Here we have adopted an arrangement where the rows of the sinogram correspond to the different projections and the columns correspond to the detector elements. While this setup is easily changeable, it depends on the arrangement of the forward operator. Some other forward operators (like the \texttt{radon.m} function in MATLAB) may use different arrangement.}

\subsection{Complementary data}
\texttt{GelPhantom\uns extra\uns frames.mat} contains an expanded version (720 projections) of the measurement data for time step 1, stored in tomographic data structures \texttt{GelPhantomFrame1\uns b2} and \texttt{GelPhantomFrame1\uns b4} for the two respective binning levels. Additionally, 1600 projections were measured of an 18th time step. These are included in the data structures \texttt{GelPhantomFrame18\uns b2} and \texttt{GelPhantomFrame18\uns b4}. Since it is not possible to downsample 1600 evenly spaced projections to the same angular sampling as was used in the primary measurements, this 18th measurement is not included with the other sinograms. But due to the high number of projections, it can provide high quality reconstructions of the final stage of the phantom and is hence included as a complementary material.

\subsection{Forward operators}
\begin{figure}[hbt]
    \centering
    \begin{tikzpicture}
    \draw[step=1.0,black,thin] (0,0) grid (4,4);
    \node at (0.5,3.5) {$x_1$};
    \node at (0.5,2.5) {$x_2$};
    \node at (0.5,1.5) {$\vdots$};
    \node at (0.5,0.5) {$x_N$};
    \node at (1.5,3.5) {$x_{N+1}$};
    \node at (1.5,2.5) {$x_{N+2}$};
    \node at (1.5,1.5) {$\vdots$};
    \node at (1.5,0.5) {$x_{2N}$};
    \node at (2.5,3.5) {$\dots$};
    \node at (2.5,2.5) {$\dots$};
    \node at (2.5,1.5) {$\ddots$};
    \node at (2.5,0.5) {$\dots$};
    \node at (3.5,0.5) {$x_{N^2}$};

    \coordinate (C) at (-4,2); 
    \draw[gray, dashed, thick] (C)-- ($(C) + (12.2,1.95)$);
    \draw[gray, dashed, thick] (C) -- ($(C) + (12.2,-1.95)$);
    
    \filldraw[black, fill=gray!30!white] (C) -- ($(C) + (0.6,0.2) $) -- ($(C) + (0.6,-0.2)$) -- cycle;
    \filldraw[black, fill=gray!30!white] (C) circle (0.45cm) node {};
    
    \draw[->, blue, thick] ($(C) + (-0.8,0)$) arc (180:215:2cm);
    \draw[blue, thick] ($(C) + (-0.8,0)$) arc (180:145:2cm) node[pos=0, left, blue] {$\Large{\theta_i}$};

    \filldraw[black, fill=blue!30!white] ($(C) + (11.8,2)$) rectangle ($(C) + (12.4,-2)$);
    \draw[xstep=0.6,ystep=0.4,black,thin] ($(C) + (11.8,2)$) grid ($(C) + (12.4,-2)$);
    
    \node at ($(C) + (12.1,1.8)$) {\scalebox{0.8}{$m_{i,1}$}};
    \node at ($(C) + (12.1,1.4)$) {\scalebox{0.8}{$m_{i,2}$}};
    \node at ($(C) + (12.1,0.2)$) {$\cdot$};
    \node at ($(C) + (12.1,-0.2)$) {$\cdot$};
    \node at ($(C) + (12.1,-0.6)$) {$\cdot$};
    \node at ($(C) + (12.1,-1.8)$) {\scalebox{0.8}{$m_{i,D}$}};
    
    \end{tikzpicture}
    \caption{Measurement model. Distances or ratios of different directions are not in scale.} \label{fig:measmodel}
\end{figure}
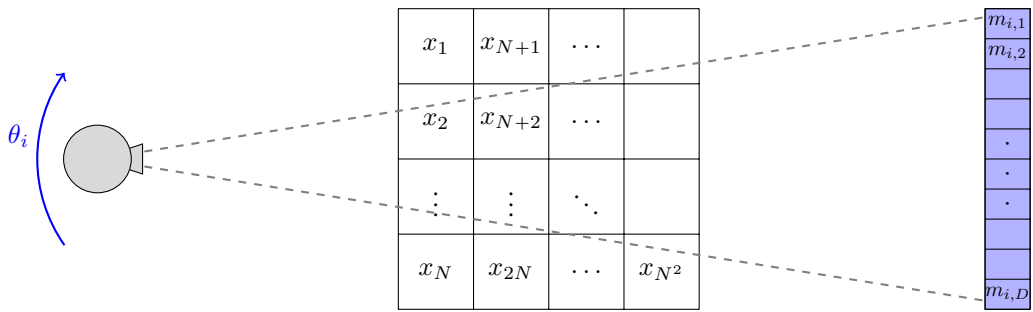

For an $N$ by $N$ square target $\vec{x}$ defined by its non-negative attenuation values $x_j,\;j\in N^2$, the mathematical model for the CT measurements can be expressed as
\[
A\vec{x(:)} = \vec{m(:)},
\]
where $A$ is the forward operator and $(:)$ indicates how the vectors $\vec{x}$ and $\vec{m}$ are considered in their column form. For $P$ projections and $D$ detector elements, $A$ can be expressed as a $PD \times N^2$ matrix with most elements zero. The values of $A$ are determined by which pixels of $\vec{x}$ are intersected by a ray from direction $\theta_i$ on the path to detector element $j$. Figure \ref{fig:measmodel} illustrates how a single projection is formed.

The metadata allows for easy construction of forward operators based on the ASTRA Toolbox. As the dimensions $P, D$ and $N$ get bigger, the operators become computationally expensive even if sparse matrices are used. Therefore the use of Spot operators is recommended. These are computationally efficient substitutes to matrices which allow for some matrix-like operations to be performed, including multiplication by scalars, vectors and matrices, transpose and Kronecker product. However these operators can not be saved into \texttt{.mat}-files and therefore have to be created when needed.

A MATLAB function for creating the operator on a CPU is provided by the HelTomo Toolbox as \texttt{create\uns ct\uns operator\uns 2d\uns fan\uns astra.m}. Computers with CUDA compatible graphics cards can also create a faster, GPU-adapted version of the operator with \texttt{create\uns ct\uns operator\uns 2d\uns fan\uns astra\uns cuda.m}. Note that since each of the time steps share the same projection geometry, as long as identical sampling of projection directions is used, the same forward operator can be used for each of the measurements. The HelTomo Toolbox also works as an interface for the ASTRA Toolbox, designed to work with X-ray data measured at the University of Helsinki.

While the reconstruction resolution $N$ can be freely chosen by the user, values of 512 and 256 corresponding to binning of 2 and 4 respectively are recommended. If the resolution is too small, outer edges of the target can be left out and if the resolution is too big, unnecessary computations need to be performed.

\subsection{Filtered back-projection reconstructions}
\begin{figure}[hbt]
    \centering
    \includegraphics[width=0.33\textwidth]{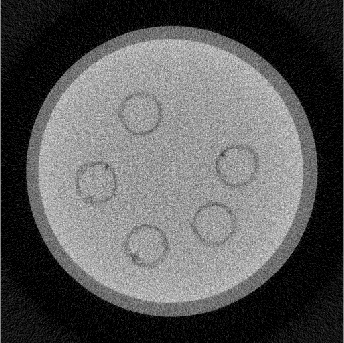}
    \includegraphics[width=0.33\textwidth]{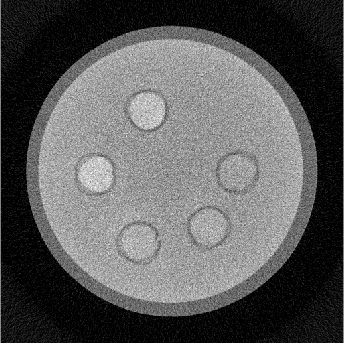}
    \includegraphics[width=0.33\textwidth]{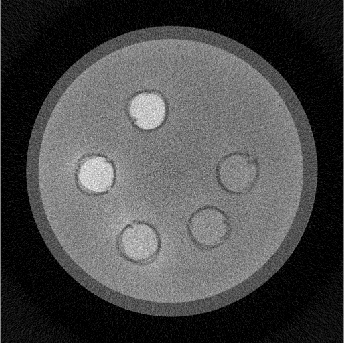}
    \caption{512 $\times$ 512 FBP reconstructions, 360 projections. Time frames 1, 7 and 12.}
    \label{fig:FBP}
\end{figure}

As with the forward operators, filtered back-projection (FBP) reconstructions can be easily computed with ASTRA via the HelTomo Toolbox utilizing the function \texttt{tomorecon\uns 2d\uns fan\uns fbp.m} on CPU, or with \texttt{tomorecon\uns 2d\uns fan\uns fbp\uns cuda.m} on a CUDA compatible GPU. Some of these reconstructions are illustrated in figure~\ref{fig:FBP}.

\section{Measurements} \label{sec:measurements}
Approximately 5 ml of contrast agent was used in total, consisting of potassium iodide (KI) diluted into 20\% sucrose water to obtain a mix with 30\% w/v of potassium iodide. In order to guarantee the direction of the diffusion, five small holes in the gel were filled with the same sucrose water before the testing began. Additional contrast agent was applied by hand between measurements to keep the rate of diffusion at a steady rate.

Initially, the iodide was applied to the centermost cavity only. After this, iodide was added to the centermost cavity and its two neighbours, and from the third application onwards to all five cavities. The aim was to simulate an increasing spot-like exposure which would spread both vertically and laterally.

Each round of measurements took 4.5 minutes and the interval between consecutive measurements was approximately 15 minutes. In total, the experiment lasted about 3 hours. Approximately halfway through the experiment no more iodide was added and from that point on only the previously applied contrast agent is diffusing into the agarose gel.

\subsection{Projector geometry}
The measurements were done in the University of Helsinki micro-CT laboratory with a Phoenix Nanotom S scanner \cite{suuronen2014bench}. An exposure time of $2\times 250$ ms, with a 250 ms pause for rotations, an X-ray tube acceleration voltage of 60 kV, and a tube current of 250~$\upmu$A were used throughout the experiment.

The distance from the X-ray source to the origin (SOD), which acts as the center of rotation, was 100.00~mm, and the distance from the origin to the detector (ODD) was 200.00~mm. The detector is a 5 megapixel CMOS flat-panel detector (Hamamatsu Photonics, Japan) with 2304 $\times$ 2304 pixels, each the size of 50 \um \ $\times$ 50 \um. Due to the scanner settings the actual projections are of size 1128 $\times$ 1152 with pixel width of 100~\um.

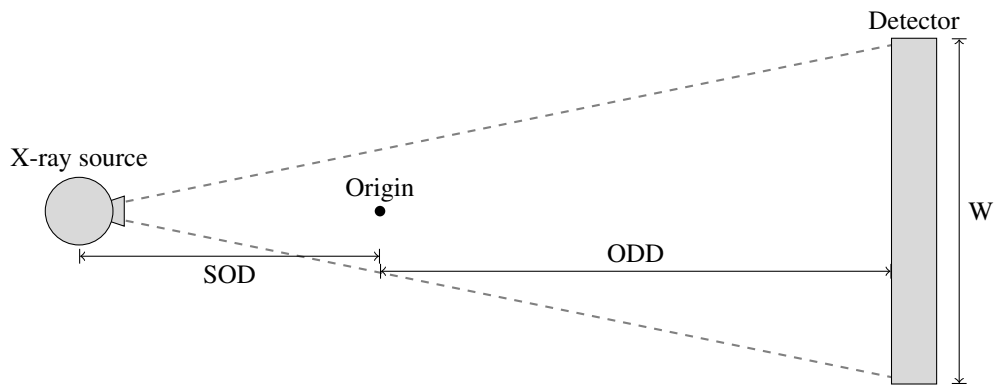
\begin{figure}[hbt]
    \centering
    \begin{tikzpicture}
    \coordinate (O) at (-2,2);
    \fill[black] (O) circle (2pt) node[above] {Origin};
    
    \coordinate (C) at (-6,2); 
    \draw[gray, dashed, thick] (C)-- ($(C) + (11,2.25)$);
    \draw[gray, dashed, thick] (C) -- ($(C) + (11,-2.25)$);
    
    \filldraw[black, fill=gray!30!white] (C) -- ($(C) + (0.6,0.2) $) -- ($(C) + (0.6,-0.2)$) -- cycle;
    \filldraw[black, fill=gray!30!white] (C) circle (0.45cm);
    \node[above] at ($(C) + (0,0.4)$) {X-ray source};
    
    \filldraw[black, fill=gray!30!white] ($(C) + (10.8,2.3)$) rectangle ($(C) + (11.4,-2.3)$);
    \node[above] at ($(C) + (11.1,2.3)$) {Detector};
    
    \draw[|<->|] (-6,1.4) -- (-2,1.4) node[midway, below] {SOD};
    \draw[|<->|] (4.8,1.2) -- (-2,1.2) node[midway, above] {ODD};
    \draw[|<->|] (5.7,4.3) -- (5.7,-0.3) node[midway, right] {W};
    
    \end{tikzpicture}
    \caption{Top-down view of the projector geometry. Distances or ratios of different directions are not necessarily to scale.} \label{fig:geometry}
\end{figure}

The SOD and ODD distances cause a geometric magnification, but this information is stored in the metadata and is taken into account by the HelTomo Toolbox and the ASTRA Toolbox. The total width (W) of the detector is 115~mm. These dimensions and the detector geometry are illustrated in figure \ref{fig:geometry}.

With this measurement setup, the target is placed on a computer controlled rotational platform which turns clockwise between projections, corresponding to the source-detector pair rotating counterclockwise around the target.

While the micro-CT scanner takes cone-beam measurements which could be used to create 3D reconstructions, this data consists of only of the center row of the detector panel, providing a fan-beam geometry, resulting in 2D reconstructions. The 3D data and additional documentation are available in \href{https://doi.org/10.5281/zenodo.7739984}{Zenodo}.

\section{Example code} \label{sec:examples}
\begin{figure}[!hbt]
    \centering
    \includegraphics[width = 0.92\linewidth]{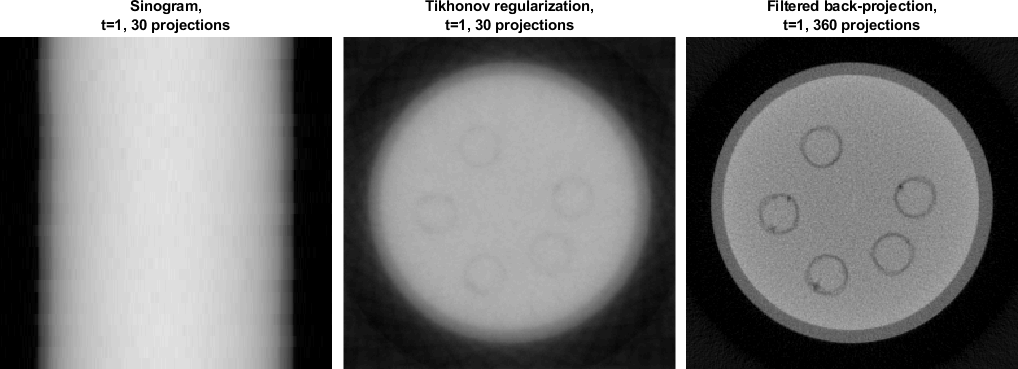} \\[1em]
    \includegraphics[width = 0.92\linewidth]{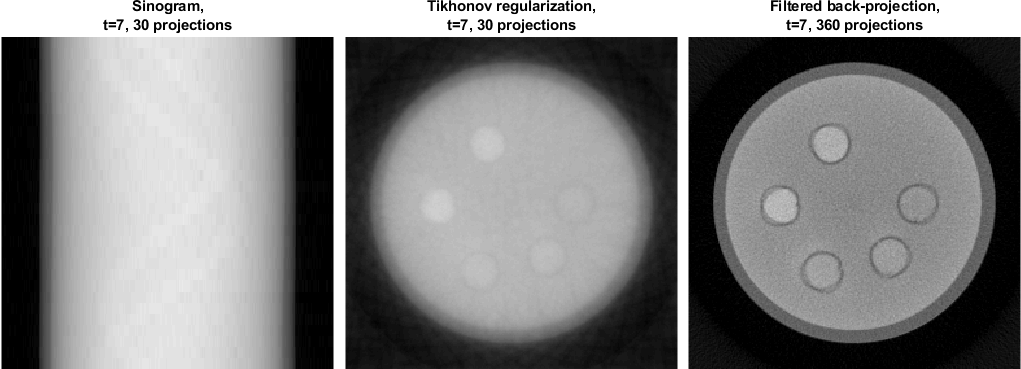} \\[1em]
    \includegraphics[width = 0.92\linewidth]{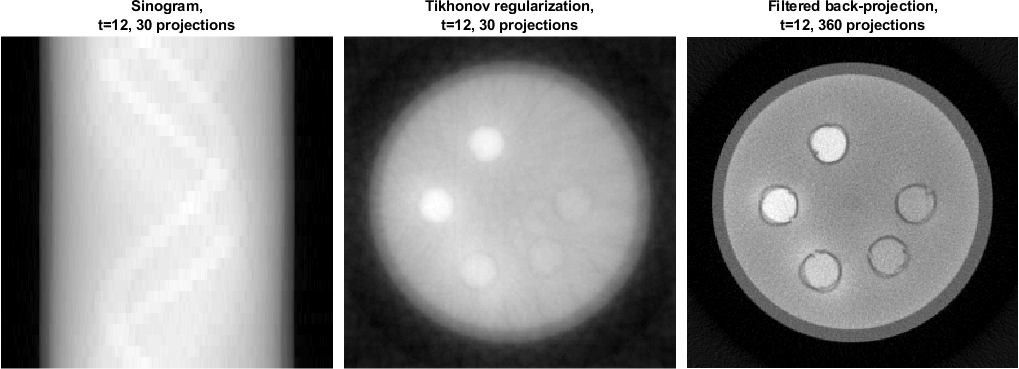}
    \caption{Reconstructions from selected time frames using \texttt{gel\uns phantom\uns example.m}.}
    \label{fig:example}
\end{figure}

The file \texttt{gel\uns phantom\uns example.m} contains a short MATLAB program to illustrate how the data set could be used. A self-contained part of it is presented below, and the resulting sinograms and reconstructions are shown in figure \ref{fig:example}. For users with CUDA compatible graphics cards, the file \texttt{gel\uns phantom\uns example\uns GPU.m} contains the same routine with GPU-adapted forward operators and FBP algorithm.

\begin{lstlisting}
% Load the data structure from the file.
load('GelPhantomData_b4.mat');
frames = [1,7,12];      % Choose desired time frames
N = 256;                % Define reconstruction size
alpha = 0.1;            % Regularization parameter
angles = [0:12:348];    % Measurement angles in degrees

% Downsample the data from the first measurement
CtData = subsample_sinogram(GelPhantomData_b4(1), angles);

 % Create the forward operator
A = create_ct_operator_2d_fan_astra(CtData, N, N);

% Normalize the operator
normA = normest(A);
A = A/normA;

% Define the forward model as a function
fun = @(x) A.'*(A*x(:))+alpha*x(:);

% Preallocate the conjugate gradient reconstruction
recn = zeros(N,N,length(frames));

% Go through the time frames
for i = 1:length(frames)
    % Pick the desired data and downsample it
    CtData = subsample_sinogram(GelPhantomData_b4(frames(i)), angles);
    % Normalize sinogram
    m = CtData.sinogram / normA;
    % Back-project the data
    b = A.'*m(:);
    % Perform conjugate gradient algorithm and save the reconstruction
    recn(:,:,i) = reshape(pcg(fun, b), N, N);
end

% Preallocate the FBP reconstruction
recnFBP = zeros(N,N,length(frames));

% Go through the time frames
for i = 1:length(frames)
    % Perform FBP on the non-downsampled data
    recnFBP(:,:,i) = tomorecon_2d_fan_fbp(GelPhantomData_b4(frames(i)),N, N);
end

% Normalize the reconstructions to interval [0,1] for viewing
recn = recn/max(recn(:));
recnFBP = recnFBP/max(recnFBP(:));

% Look at the results and the corresponding sinograms
for i = 1:length(frames)
    figure(i);
    hold on

    subplot(1,3,1)
    imagesc(GelPhantomData_b4(frames(i)).sinogram(1:12:349,:),[])
    colormap gray
    axis square
    axis off
    title({'Sinogram,'; strcat('t=', num2str(frames(i)), ', 30 projections')})
    
    subplot(1,3,2)
    imagesc(recn(:,:,i),[0,1])
    colormap gray
    axis square
    axis off
    title({'Tikhonov regularization,'; strcat('t=', num2str(frames(i)), ...
            ', 30 projections')})
    
    subplot(1,3,3)
    imagesc(recnFBP(:,:,i),[0,1])
    colormap gray
    axis square
    axis off
    title({'Filtered back-projection,'; strcat('t=', num2str(frames(i)), ...
            ', 360 projections')})
    hold off
end
\end{lstlisting}
\vspace{-10pt}

\bibliographystyle{unsrt}  
\bibliography{references}

\end{document}